# Pass-Join: A Partition-based Method for Similarity Joins

Guoliang Li    Dong Deng    Jiannan Wang    Jianhua Feng
Department of Computer Science and Technology, Tsinghua University, Beijing 100084, China.
liguoliang@tsinghua.edu.cn, dd11@mails.thu.edu.cn, wjn08@mails.thu.edu.cn, fengjh@tsinghua.edu.cn

## ABSTRACT

As an essential operation in data cleaning, the similarity join has attracted considerable attention from the database community. In this paper, we study string similarity joins with edit-distance constraints, which find *similar* string pairs from two large sets of strings whose edit distance is within a given threshold. Existing algorithms are efficient either for short strings or for long strings, and there is no algorithm that can efficiently and adaptively support both short strings and long strings. To address this problem, we propose a partition-based method called Pass-Join. Pass-Join partitions a string into a set of segments and creates inverted indices for the segments. Then for each string, Pass-Join selects some of its substrings and uses the selected substrings to find candidate pairs using the inverted indices. We devise efficient techniques to select the substrings and prove that our method can minimize the number of selected substrings. We develop novel pruning techniques to efficiently verify the candidate pairs. Experimental results show that our algorithms are efficient for both short strings and long strings, and outperform state-of-the-art methods on real datasets.

## 1. INTRODUCTION

A string similarity join between two sets of strings finds all *similar* string pairs from the two sets. For example, consider two sets of strings {vldb, sigmod, ... } and {pvldb, icde, ... }. We want to find all similar pairs, e.g., ⟨vldb, pvldb⟩. Many similarity functions have been proposed to quantify the similarity between two strings, such as Jaccard similarity, Cosine similarity, and edit distance. In this paper, we study string similarity joins with edit-distance constraints, which, given two sets of strings, find all *similar* string pairs from the two sets, such that the edit distance between each string pair is within a given threshold. The string similarity join is an essential operation in many applications, such as data integration and cleaning, near duplicate object detection and elimination, and collaborative filtering.

Existing methods to address this problem can be broadly classified into two categories. The first one uses a *filter-and-refine* framework, such as Part-Enum [2], All-Pairs-Ed [3], ED-Join [23]. In the *filter* step, they generate signatures for each string and use the signatures to generate candidate pairs. In the *refine* step, they verify the candidate pairs to generate the final results. However, these approaches are inefficient for the datasets with short strings (e.g., person names and locations) [20]. The main reason is that they cannot select high-quality signatures for short strings and will generate large numbers of candidates which need to be further verified. The second method (Trie-Join [20]) adopts a trie-based framework, which uses a trie structure to share prefixes and utilizes prefix pruning to improve the performance. However Trie-Join is inefficient for long strings (e.g., paper titles and abstracts). This is because long strings have a small number of shared prefixes.

If a system wants to support both short strings and long strings, we have to implement and maintain two separate codes, and tune many parameters to select the best method. To alleviate this problem, it calls for an adaptive method which can efficiently support both short strings and long strings. In this paper we propose a partition-based method to address this problem. We devise a partition scheme to partition a string into a set of segments and prove that if a string $s$ is similar to string $r$, $s$ must have a substring which matches a segment of $r$. Based on this observation, we propose a *p*artition-based framework for *s*tring *s*imilarity *joins*, called Pass-Join. Pass-Join creates inverted indices for the segments. For each string $s$, we select some of its substrings, and search for the selected substrings in the inverted indices. If a selected substring appears in the inverted index, each string $r$ on the inverted list of this substring (i.e., $r$ contains the substring) may be similar to $s$, and we take $r$ and $s$ as a candidate pair. Next we verify the pair to generate the final answers. We develop effective techniques to select substrings and prove that our method can minimize the number of selected substrings. We devise novel pruning techniques to verify candidate pairs. To summarize, we make the following contributions.

(1) We devise a partition scheme to partition strings into a set of segments. Using the partition scheme, we propose a partition-based framework to facilitate similarity joins.

(2) We develop novel techniques to select substrings and use them to generate candidate pairs. We prove that our method can minimize the number of selected substrings.

(3) We propose an extension-based method to efficiently verify a candidate pair, and develop pruning techniques and

early-termination techniques to improve the performance.

(4) We have conducted an extensive set of experiments. Experimental results show that our algorithms are very efficient for both short strings and long strings, and outperform state-of-the-art methods on real datasets.

The rest of this paper is organized as follows. We formalize our problem in Section 2. Section 3 introduces our partition-based framework. We propose to effectively select substrings in Section 4 and develop novel techniques to efficiently verify candidates in Section 5. Experimental results are provided in Section 6. We review related work in Section 7 and make a conclusion in Section 8.

## 2. PROBLEM FORMULATION

Given two collections of strings, a similarity join finds all *similar* string pairs from the two collections. In this paper, we use edit distance to quantify the similarity between two strings. Formally, the edit distance between two strings $r$ and $s$, denoted by $\text{ED}(r,s)$, is the minimum number of single-character edit operations (i.e., insertion, deletion, and substitution) needed to transform $r$ to $s$. For example, $\text{ED}(\text{"kaushic chaduri"}, \text{"kaushuk chadhui"}) = 4$.

In this paper two strings are *similar* if their edit distance is not larger than a specified edit-distance threshold $\tau$. We formalize the problem of string similarity joins as follows.

DEFINITION 1 (STRING SIMILARITY JOINS). *Given two sets of strings $\mathcal{R}$ and $\mathcal{S}$ and an edit-distance threshold $\tau$, a similarity join finds all similar string pairs $\langle r,s \rangle \in \mathcal{R} \times \mathcal{S}$ such that $\text{ED}(r,s) \leq \tau$.*

Without loss of generality, we focus on self join in this paper, that is $\mathcal{R} = \mathcal{S}$. We will discuss how to join two distinct sets ($\mathcal{R} \neq \mathcal{S}$) in Section 3.

For example, consider the strings in Table 1(a). Suppose threshold $\tau=3$. $\langle$"kaushik chakrab", "caushik chakrabar"$\rangle$ is a similar pair as their edit distance is not larger than $\tau$.

Table 1: A set of strings

| (a) Strings |     | (b) Sorted strings |        |
|-------------|-----|--------------------|--------|
| Strings     | ID  | Strings            | Length |
| avataresha  | $s_1$ | vankatesh        | 9      |
| caushik chakrabar | $s_2$ | avataresha  | 10     |
| kaushic chaduri | $s_3$ | kaushic chaduri | 15   |
| kaushik chakrab | $s_4$ | kaushik chakrab | 15   |
| kaushuk chadhui | $s_5$ | kaushuk chadhui | 15   |
| vankatesh   | $s_6$ | caushik chakrabar | 17   |

## 3. PARTITION-BASED SIMILARITY JOINS

We first introduce a partition scheme to partition a string into several disjoint segments (Section 3.1), and then propose a partition-based framework (Section 3.2).

### 3.1 Partition Scheme

Given a string $s$, we partition it into $\tau + 1$ disjoint segments, and the length of each segment is not smaller than one[*]. For example, consider string $s_1=$"vankatesh". Suppose $\tau = 3$. We have multiple ways to partition $s_1$ into $\tau + 1 = 4$ segments, such as {"va","nk","at", "esh"}.

Consider two strings $r$ and $s$. If $s$ has no substring that matches a segment of $r$, $s$ cannot be similar to $r$ based on the pigeonhole principle as stated in Lemma 1. Due to space constraints, we refer readers to our technical report [16]. In other words, if $s$ is similar to $r$, $s$ must contain a substring matching a segment of $r$. For example, consider strings in Table 2. Suppose $\tau=3$. $s_1=$"vankatesh" has four segments {"va", "nk", "at", "esh"}. As $s_3, s_4, s_5, s_6$ have no substrings matching segments of $s_1$, they are not similar to $s_1$.

LEMMA 1. *Given a string $r$ with $\tau + 1$ segments and a string $s$, if $s$ is similar to $r$ within threshold $\tau$, $s$ must contain a substring which matches a segment of $r$.*

Given a string, there could be many strategies to partition the string into $\tau+1$ segments. A good partition strategy can reduce the number of candidate pairs and thus improve the performance. Intuitively, the shorter a segment of $r$ is, the higher probability the segment appears in other strings, and the more strings will be taken as $r$'s candidates, thus the pruning power is lower. Based on this observation, we do not want to keep short segments in the partition. In other words, each segment should have nearly the same length. Accordingly we propose an *even-partition* scheme. Consider a string $s$ with length $|s|$. In even partition, each segment has a length of $\lfloor \frac{|s|}{\tau+1} \rfloor$ or $\lceil \frac{|s|}{\tau+1} \rceil$, thus the maximal length difference between two segments is 1. Let $k = |s| - \lfloor \frac{|s|}{\tau+1} \rfloor * (\tau + 1)$. In even partition, the last $k$ segments have length $\lceil \frac{|s|}{\tau+1} \rceil$, and the first $\tau + 1 - k$ ones have length $\lfloor \frac{|s|}{\tau+1} \rfloor$. For example, consider $s_1=$"vankatesh" and suppose $\tau = 3$. We have $k = 1$. $s_1$ has four segments {"va","nk","at", "esh"}.

Although we can devise other partition schemes, it is time consuming to select a good partition strategy. Note that the time for selecting a partition strategy should be included in the similarity join time. In this paper we use the even-partition scheme and leave the problem of selecting good partition strategies as a future work. Note that our proposed techniques can be extended to other partition strategies.

### 3.2 Partition-based Framework

We have an observation that if a strings $s$ does not have a substring that matches a segment of $r$, we can prune the pair $\langle s, r \rangle$. We can use this feature to prune large numbers of dissimilar pairs. To this end, we propose a *partition-based framework for string similarity joins*, called PASS-JOIN. Figure 2 illustrates our framework.

For ease of presentation, we first introduce some notations. Let $\mathcal{S}_l$ denote the set of strings with length $l$ and $\mathcal{S}_l^i$ denote the set of the $i$-th segments of strings in $\mathcal{S}_l$. We build an inverted index for each $\mathcal{S}_l^i$, denoted by $\mathcal{L}_l^i$. Given an $i$-th segment $w$, let $\mathcal{L}_l^i(w)$ denote the inverted list of segment $w$, i.e., the set of strings whose $i$-th segments are $w$. PASS-JOIN uses the inverted indices to do similarity joins as follows.

PASS-JOIN first sorts strings based on their lengths in ascending order. For the strings with the same length, it sorts them in alphabetical order. Then PASS-JOIN visits strings in order. Consider the current string $s$ with length $|s|$. PASS-JOIN finds $s$'s similar strings among the visited strings using the inverted indices. To efficiently find such strings, we create indices only for visited strings to avoid enumerating a string pair twice. Based on length filtering [7], we check whether the strings in $\mathcal{L}_l^i$ ($|s| - \tau \leq l \leq |s|, 1 \leq i \leq \tau + 1$) are similar to $s$. Without loss of generality, consider inverted index $\mathcal{L}_l^i$. PASS-JOIN finds $s$'s similar strings in $\mathcal{L}_l^i$ as follows.

- SUBSTRING SELECTION: If $s$ is similar to a string in $\mathcal{L}_l^i$, $s$ should contain a substring which matches a segment in $\mathcal{L}_l^i$. A straightforward method enumerates all of $s$'s substrings, and for each substring checks whether it appears in $\mathcal{L}_l^i$. Actually we do not need to consider all substrings of $s$. Instead we only select some substrings (denoted by $\mathcal{W}(s, \mathcal{L}_l^i)$) and use the selected substrings to find similar pairs. We discuss how to generate $\mathcal{W}(s, \mathcal{L}_l^i)$ in Section 4. For each selected substring $w \in \mathcal{W}(s, \mathcal{L}_l^i)$, we check whether it appears in $\mathcal{L}_l^i$. If so, for each $r \in \mathcal{L}_l^i(w)$, $\langle r, s \rangle$ is a candidate pair.

---

[*]The length of string $s(|s|)$ should be larger than $\tau$, i.e., $|s| \geq \tau + 1$.

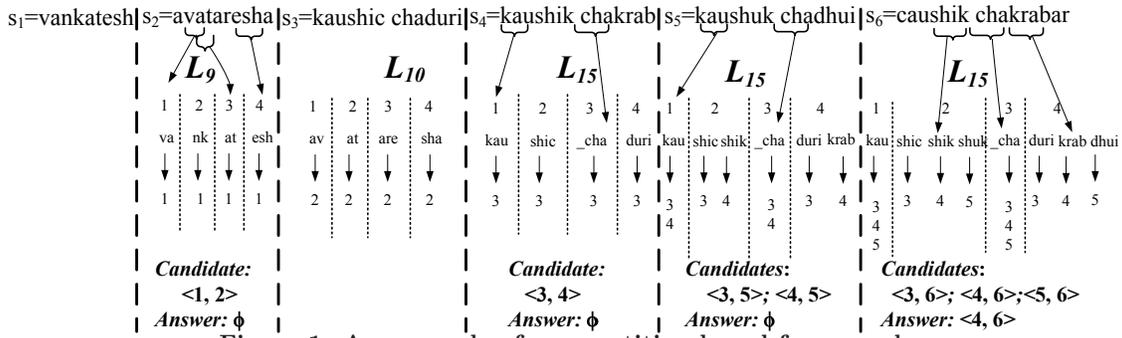

Figure 1: An example of our partition-based framework

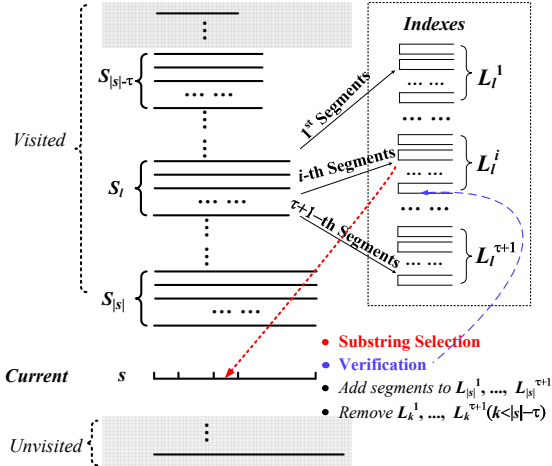

Figure 2: Partition-based framework

- VERIFICATION: To verify whether a candidate pair $\langle r, s \rangle$ is an answer, a straightforward method computes their real edit distance. However this method is rather expensive, and we develop effective techniques to do efficient verification in Section 5.

After finding similar strings for $s$, we partition $s$ into $\tau + 1$ segments and insert the segments into inverted index $\mathcal{L}_{|s|}^i (1 \leq i \leq \tau+1)$. Then we visit strings after $s$ and iteratively we can find all similar pairs. Note that we can remove the inverted index $\mathcal{L}_k^i$ for $k < |s|-\tau$. Thus we maintain at most $(\tau+1)^2$ inverted indices $\mathcal{L}_l^i$ for $|s|-\tau \leq l \leq |s|$ and $1 \leq i \leq \tau+1$.

To join two distinct sets $\mathcal{R}$ and $\mathcal{S}$, we first sort the strings in the two sets respectively. Then we index the segments of strings in a set, e.g., $\mathcal{S}$. Next we visit strings of $\mathcal{R}$ in order. For each string $r \in \mathcal{R}$ with length $|r|$, we use the inverted indices of strings in $\mathcal{S}$ with lengths between $[|r|-\tau, |r|+\tau]$ to find similar pairs. We can remove the indices for strings with lengths smaller than $|r|-\tau$. In this paper we focus on the case that the index can be fit in the memory. We leave dealing with a very large dataset as a future work.

For example, consider strings in Table 1. Suppose $\tau = 3$. We find similar pairs as follows (Figure 1). For the first string $s_1 = $ "vankatesh", we partition it into $\tau + 1$ segments and insert the segments into the inverted indices for strings with length 9, i.e., $\mathcal{L}_9^1, \mathcal{L}_9^2, \mathcal{L}_9^3,$ and $\mathcal{L}_9^4$. Next for $s_2 = $ "avataresha", we enumerate its substrings and check whether each substring appears in $\mathcal{L}_{|s_2|-\tau}^i, \cdots, \mathcal{L}_{|s_2|}^i (1 \leq i \leq \tau+1)$. Here we find "va" in $\mathcal{L}_9^1$, "at" in $\mathcal{L}_9^3$, and "esh" in $\mathcal{L}_9^4$. For segment "va", as $\mathcal{L}_9^1(\text{va}) = \{s_1\}$. The pair $\langle s_2, s_1 \rangle$ is a candidate pair. We verify the pair and it is not an answer as the edit distance is larger than $\tau$. Next we partition $s_2$ into four segments and insert them into $\mathcal{L}_{|s_2|}^1, \mathcal{L}_{|s_2|}^2, \mathcal{L}_{|s_2|}^3, \mathcal{L}_{|s_2|}^4$. Similarly we repeat the above steps and find all similar pairs.

We give the pseudo-code of our algorithm in Figure 3. PASS-JOIN sorts strings first by length and then in alphabetical order (line 2). Then, PASS-JOIN visits each string in sorted order (line 3). For each inverted index $\mathcal{L}_l^i (|s|-\tau \leq l \leq |s|, 1 \leq i \leq \tau+1)$, PASS-JOIN selects the substrings of $s$ (line 4) and checks whether each selected substring $w$ is in $\mathcal{L}_l^i$ (line 5). If yes, for any string $r$ in the inverted list of $w$

---

**Algorithm 1**: PASS-JOIN $(\mathcal{S}, \tau)$

**Input**: $\mathcal{S}$: A collection of strings
$\tau$: A given edit-distance threshold
**Output**: $\mathcal{A} = \{(s \in \mathcal{S}, r \in \mathcal{S}) \mid \text{ED}(s, r) \leq \tau\}$

1 **begin**
2    Sort $\mathcal{S}$ first by string length and second in alphabetical order;
3    **for** $s \in \mathcal{S}$ **do**
4      **for** $\mathcal{L}_l^i$ ($|s|-\tau \leq l \leq |s|, 1 \leq i \leq \tau+1$) **do**
5        $\mathcal{W}(s, \mathcal{L}_l^i) = $ SUBSTRINGSELECTION$(s, \mathcal{L}_l^i)$;
6        **for** $w \in \mathcal{W}(s, \mathcal{L}_l^i)$ **do**
7          **if** $w$ is in $\mathcal{L}_l^i$ **then**
           VERIFICATION$(s, \mathcal{L}_l^i(w), \tau)$;
8    Partition $s$ and add its segments into $\mathcal{L}_{|s|}^i$;
9 **end**

**Function** SUBSTRINGSELECTION$(s, \mathcal{L}_l^i)$

**Input**: $s$: A string; $\mathcal{L}_l^i$: Inverted index
**Output**: $\mathcal{W}(s, \mathcal{L}_l^i)$: Selected substrings

1 **begin**
2    $\mathcal{W}(s, \mathcal{L}_l^i) = \{w \mid w \text{ is a substring of } s\}$;
3 **end**

**Function** VERIFICATION$(s, \mathcal{L}_l^i(w), \tau)$

**Input**: $s$: A string; $\mathcal{L}_l^i(w)$: Inverted list; $\tau$: Threshold
**Output**: $\mathcal{A} = \{(s \in \mathcal{S}, r \in \mathcal{S}) \mid \text{ED}(s, r) \leq \tau\}$

1 **begin**
2    **for** $r \in \mathcal{L}_l^i(w)$ **do**
3      **if** ED$(s, r) \leq \tau$ **then**    $\mathcal{A} \leftarrow \langle s, r \rangle$;
4 **end**

Figure 3: Pass-Join algorithm

in $\mathcal{L}_l^i$, i.e., $\mathcal{L}_l^i(w)$, the string pair $\langle r, s \rangle$ is a candidate pair. PASS-JOIN verifies the pair (line 7). Finally, PASS-JOIN partitions $s$ into $\tau + 1$ segments, and inserts the segments into the inverted index $\mathcal{L}_{|s|}^i (1 \leq i \leq \tau + 1)$ (line 8). Here function SUBSTRINGSELECTION selects all substrings and function VERIFICATION computes the real edit distance of two strings to verify the candidates using dynamic-programming algorithm. To improve the performance, we propose effective techniques to improve the substring-selection step in Section 4 and the verification step in Section 5.

**Complexity:** We first analyze the space complexity. Our indexing structure includes segments and inverted lists of segments. We first give the space complexity of segments. For each string in $\mathcal{S}_l$ we generate $\tau + 1$ segments. Thus the number of segments is at most $(\tau+1) \times |\mathcal{S}_l|$, where $|\mathcal{S}_l|$ is the number strings in $\mathcal{S}_l$. As we can use an integer to encode a segment, the space complexity of segments is

$$\mathcal{O}\Big( \max_{l_{min} \leq j \leq l_{max}} \sum_{l=j-\tau}^{j} (\tau+1) \times |\mathcal{S}_l| \Big),$$

where $l_{min}$ and $l_{max}$ respectively denote the minimal string length and the maximal string length.

Next we give the complexity of inverted lists. For each string in $\mathcal{S}_l$, as the $i$-th segment of the string corresponds to an element in $\mathcal{L}_l^i$, $|\mathcal{S}_l| = |\mathcal{L}_l^i|$. The space complexity of inverted lists(i.e., the sum of the lengths of inverted lists) is

$$\mathcal{O}\Big( \max_{l_{min} \leq j \leq l_{max}} \sum_{l=j-\tau}^{j} \sum_{i=1}^{\tau+1} |\mathcal{L}_l^i| = \max_{l_{min} \leq j \leq l_{max}} \sum_{l=j-\tau}^{j} (\tau+1) \times |\mathcal{S}_l| \Big).$$

Then we give the time complexity. To sort the strings, we can first group the strings based on lengths and then sort strings in each group. Thus the sort complexity is $\mathcal{O}(\sum_{l_{min} \leq l \leq l_{max}} |\mathcal{S}_l| log(|\mathcal{S}_l|))$. For each string $s$, we select its substring set $\mathcal{W}(s, \mathcal{L}_l^i)$ for $|s| - \tau \leq l \leq |s|, 1 \leq i \leq \tau + 1$. The selection complexity is $\mathcal{O}\Big( \sum_{s \in \mathcal{S}} \sum_{l=|s|-\tau}^{|s|} \sum_{i=1}^{\tau+1} \mathcal{X}(s, \mathcal{L}_l^i) \Big)$, where $\mathcal{X}(s, \mathcal{L}_l^i)$ is the selection time complexity for $\mathcal{W}(s, \mathcal{L}_l^i)$, which is $\mathcal{O}(\tau)$ (Section 4). The selection complexity is $\mathcal{O}(\tau^3 |\mathcal{S}|)$. For each substring $w \in \mathcal{W}(s, \mathcal{L}_l^i)$, we verify whether strings in $\mathcal{L}_l^i(w)$ are similar to $s$. The verification complexity is $\mathcal{O}\Big( \sum_{s \in \mathcal{S}} \sum_{l=|s|-\tau}^{|s|} \sum_{i=1}^{\tau+1} \sum_{w \in \mathcal{W}(s, \mathcal{L}_l^i)} \sum_{r \in \mathcal{L}_l^i(w)} \mathcal{V}(s, r) \Big)$, where $\mathcal{V}(s, r)$ is the complexity for verifying $\langle s, r \rangle$, which is $\mathcal{O}(\tau * \min(|s|, |r|))$(Section 5). In the paper we propose to reduce the size of $\mathcal{W}(s, \mathcal{L}_l^i)$ and improve the verification cost $\mathcal{V}(s, r)$.

## 4. IMPROVING SUBSTRING SELECTION

For any string $s \in \mathcal{S}$ and a length $l$ ($|s| - \tau \leq l \leq |s|$), we select a substring set $\mathcal{W}(s, l) = \cup_{i=1}^{\tau+1} \mathcal{W}(s, \mathcal{L}_l^i)$ of $s$ and use substrings in $\mathcal{W}(s, l)$ to find the candidates of $s$. We need to guarantee completeness of the method using $\mathcal{W}(s, l)$ to find candidate pairs. That is any similar pair must be found as a candidate pair. Next we give the formal definition.

DEFINITION 2 (COMPLETENESS). *A substring selection method satisfies completeness, if for any string $s$ and a length $l(|s| - \tau \leq l \leq |s|)$, $\forall r$ with length $l$ which is similar to $s$ and visited before $s$, $r$ must have an $i$-th segment $r_m$ which matches a substring $s_m \in \mathcal{W}(s, \mathcal{L}_l^i)$ where $1 \leq i \leq \tau + 1$.*

A straightforward method is to add all substrings of $s$ into $\mathcal{W}(s, l)$. As $s$ has $|s| - i + 1$ substrings with length $i$, the total number of $s$'s substrings is $\sum_{i=1}^{|s|}(|s|-i+1) = \frac{|s|*(|s|+1)}{2}$. For long strings, there are large numbers of substrings and it is rather expensive to enumerate all substrings.

Intuitively, the smaller size of $\mathcal{W}(s, l)$, the higher performance. Thus we want to find substring sets with smaller sizes. In this section, we propose several methods to select the substring set $\mathcal{W}(s, l)$. As $\mathcal{W}(s, l) = \cup_{i=1}^{\tau+1} \mathcal{W}(s, \mathcal{L}_l^i)$ and we want to use index $\mathcal{L}_l^i$ to do efficient filtering, next we focus on how to generate $\mathcal{W}(s, \mathcal{L}_l^i)$ for $\mathcal{L}_l^i$.

**Length-based Method:** As segments in $\mathcal{L}_l^i$ have the same length, denoted by $l_i$, the length-based method selects all substrings of $s$ with length $l_i$, denoted by $\mathcal{W}_\ell(s, \mathcal{L}_l^i)$. Let $\mathcal{W}_\ell(s, l) = \cup_{i=1}^{\tau+1} \mathcal{W}_\ell(s, \mathcal{L}_l^i)$. The length-based method satisfies completeness, as it selects all substrings with length $l_i$. The size of $\mathcal{W}_\ell(s, \mathcal{L}_l^i)$ is $|\mathcal{W}_\ell(s, \mathcal{L}_l^i)| = |s| - l_i + 1$, and the number of selected substrings is $|\mathcal{W}_\ell(s, l)| = (\tau+1)(|s|+1) - l$.

**Shift-based Method:** However the length-based method does not consider the positions of segments. To address this problem, Wang et al. [22] proposed a shift-based method to address the entity identification problem. We can extend their method to support our problem as follows. As segments in $\mathcal{L}_l^i$ have the same length, they have the same start position, denoted by $p_i$, where $p_1 = 1$ and $p_i = p_1 + \sum_{k=1}^{i-1} l_k$ for $i > 1$. The shift-based method selects $s$'s substrings with start positions in $[p_i - \tau, p_i + \tau]$ and with length $l_i$, denoted by $\mathcal{W}_f(s, \mathcal{L}_l^i)$. Let $\mathcal{W}_f(s, l) = \cup_{i=1}^{\tau+1} \mathcal{W}_f(s, \mathcal{L}_l^i)$. The size of $\mathcal{W}_f(s, \mathcal{L}_l^i)$ is $|\mathcal{W}_f(s, \mathcal{L}_l^i)| = 2\tau + 1$. The number of selected substrings is $|\mathcal{W}_f(s, l)| = (\tau+1)(2\tau+1)$.

The basic idea behind the method is as follows. Suppose a substring $s_m$ of $s$ with start position smaller than $p_i - \tau$ or larger than $p_i + \tau$ matches a segment in $\mathcal{L}_l^i$. Consider a string $r \in \mathcal{L}_l^i(s_m)$. We can partition $s(r)$ into three parts: the matching part $s_m(r_m)$, the left part before the matching part $s_l(r_l)$, and the right part after the matching part $s_r(r_r)$. As the start position of $r_m$ is $p_i$ and the start position of $s_m$ is smaller than $p_i - \tau$ or larger than $p_i + \tau$, the length difference between $s_l$ and $r_l$ must be larger than $\tau$. If we align the two strings by matching $s_m$ and $r_m$ (i.e., transforming $r_l$ to $s_l$, matching $r_m$ with $s_m$, and transforming $r_r$ to $s_r$), they will not be similar, thus we can prune substring $s_m$. Hence the shift-based method satisfies completeness.

However, the shift-based method still involves many unnecessary substrings. For example, consider two strings $s_1$ = "vankatesh" and $s_2$ = "avataresha". Suppose $\tau = 3$ and "vankatesh" is partitioned into four segments {va, nk, at, esh}. $s_2$ = "avataresha" contains a substring "at" which matches the third segment in "vank<u>at</u>esh", the shift-based method will select it as a substring. However we can prune it and the reason is as follows. Suppose we partition the two strings into three parts based on the matching segment. For instance, we partition "vankatesh" into {"vank", "at", "esh"}, and "avataresha" into {"av", "at", "aresha"}. Obviously the minimal edit distance (length difference) between the left parts ("vank" and "av") is 2 and the minimal edit distance (length difference) between the right parts ("esh" and "aresha") is 3. Thus if we align the two strings using the matching segment "at", they will not be similar. In this way, we can prune the substring "at".

### 4.1 Position-aware Substring Selection

Notice that all the segments in $\mathcal{L}_l^i$ have the same length $l_i$ and the same start position $p_i$. Without loss of generality,

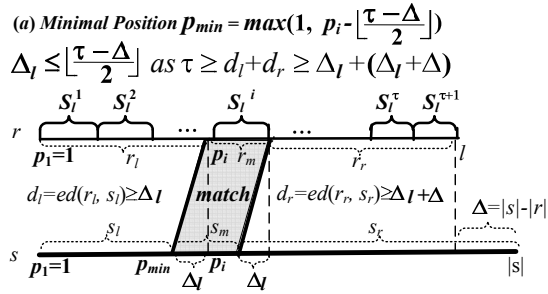

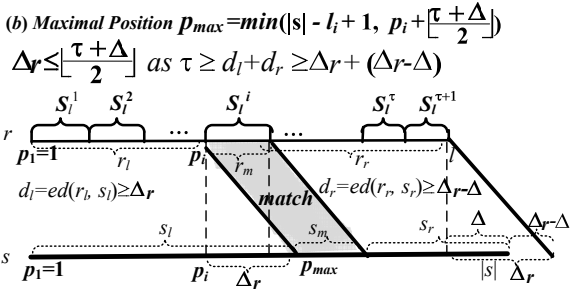

**Figure 4: Position-based substring selection**

we consider a segment $r_m \in \mathcal{L}_l^i$. Moreover, all the strings in inverted list $\mathcal{L}_l^i(r_m)$ have the same length $l$ ($l \leq |s|$), and we consider a string $r$ that contains segment $r_m$. Suppose $s$ has a substring $s_m$ which matches $r_m$. Next we give the possible start positions of $s_m$. We still partition $s(r)$ into three parts: the matching part $s_m(r_m)$, the left part $s_l(r_l)$, and the right part $s_r(r_r)$. If we align $r$ and $s$ by matching $r_m = s_m$, that is we transform $r$ to $s$ by first transforming $r_l$ to $s_l$ with $d_l = \text{ED}(r_l, s_l)$ edit operations, then matching $r_m$ with $s_m$, and finally transforming $r_r$ to $s_r$ with $d_r = \text{ED}(r_r, s_r)$ edit operations, the total transformation distance is $d_l + d_r$. If $s$ is similar to $r$, $d_l + d_r \leq \tau$. Based on this observation, we give $s_m$'s minimal start position ($p_{min}$) and the maximal start position ($p_{max}$) as illustrated in Figure 4.

**Minimal Start Position:** Suppose the start position of $s_m$, denoted by $p$, is not larger than $p_i$. Let $\triangle = |s| - |r|$ and $\triangle_l = p_i - p$. We have $d_l = \text{ED}(r_l, s_l) \geq \triangle_l$ and $d_r = \text{ED}(r_r, s_r) \geq \triangle_l + \triangle$, as illustrated in Figure 4(a). If $s$ is similar to $r$ (or **any** string in $\mathcal{L}_l^i(r_m)$), we have

$$\triangle_l + (\triangle_l + \triangle) \leq d_l + d_r \leq \tau.$$

That is $\triangle_l \leq \lfloor \frac{\tau - \triangle}{2} \rfloor$ and $p = p_i - \triangle_l \geq p_i - \lfloor \frac{\tau - \triangle}{2} \rfloor$. Thus $p_{min} \geq p_i - \lfloor \frac{\tau - \triangle}{2} \rfloor$. As $p_{min} \geq 1$, $p_{min} = \max(1, p_i - \lfloor \frac{\tau - \triangle}{2} \rfloor)$.

**Maximal Start Position:** Suppose the start position of $s_m$, $p$, is larger than $p_i$. Let $\triangle = |s| - |r|$ and $\triangle_r = p - p_i$. We have $d_l = \text{ED}(r_l, s_l) \geq \triangle_r$ and $d_r = \text{ED}(r_r, s_r) \geq |\triangle_r - \triangle|$ as illustrated in Figure 4(b). If $\triangle_r \leq \triangle$, $d_r \geq \triangle - \triangle_r$. Thus $\triangle = \triangle_r + (\triangle - \triangle_r) \leq d_l + d_r \leq \tau$, and in this case, the maximal value of $\triangle_r$ is $\triangle$; otherwise if $\triangle_r > \triangle$, $d_r \geq \triangle_r - \triangle$. If $s$ is similar to $r$ (or **any** string in $\mathcal{L}_l^i(r_m)$), we have

$$\triangle_r + (\triangle_r - \triangle) \leq d_l + d_r \leq \tau.$$

That is $\triangle_r \leq \lfloor \frac{\tau + \triangle}{2} \rfloor$, and $p = p_i + \triangle_r \leq p_i + \lfloor \frac{\tau + \triangle}{2} \rfloor$. Thus $p_{max} \leq p_i + \lfloor \frac{\tau + \triangle}{2} \rfloor$. As the segment length is $l_i$, based on the boundary, we have $p_{max} \leq |s| - l_i + 1$. Thus $p_{max} = \min(|s| - l_i + 1, p_i + \lfloor \frac{\tau + \triangle}{2} \rfloor)$.

For example, consider string $r =$ "`vankatesh`". Suppose $\tau = 3$ and "`vankatesh`" is partitioned into four segments, {`va`, `nk`, `at`, `esh`}. For string $s =$ "`avataresha`", we have $\triangle = |s| - |r| = 1$. $\triangle_l \leq \lfloor \frac{\tau - \triangle}{2} \rfloor = 1$ and $\triangle_r \leq \lfloor \frac{\tau + \triangle}{2} \rfloor = 2$. For the first segment "`va`", $p_1 = 1$. $p_{min} = \max(1, p_1 - \lfloor \frac{\tau - \triangle}{2} \rfloor) = 1$ and $p_{max} = 1 + \lfloor \frac{\tau + \triangle}{2} \rfloor = 3$. Thus we only need to enumerate the following substrings "`av`", "`va`", "`at`" for the first segment. Similarly, we need to enumerate substrings "`va`", "`at`", "`ta`", "`ar`" for the second segment, "`ta`", "`ar`", "`re`", "`es`" for the third segment, and "`res`", "`esh`", "`sha`" for the fourth segment. We see that the position-based method can reduce many unnecessary substrings over the shift-based method (reducing the number from 28 to 14).

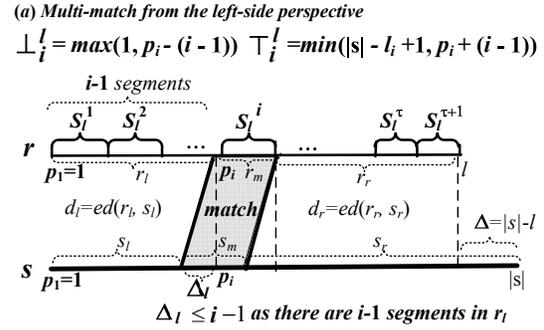

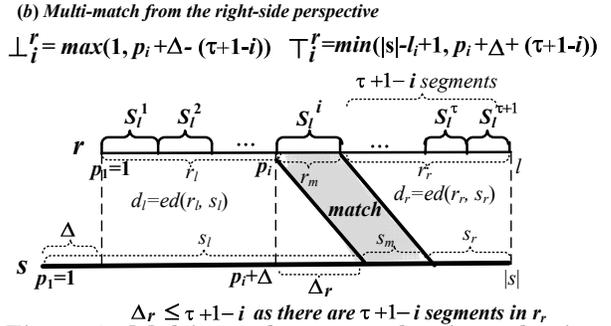

**Figure 5: Multi-match-aware substring selection**

For $\mathcal{L}_l^i$, the position-aware method selects substrings with start positions in $[p_{min}, p_{max}]$ and with length $l_i$, denoted by $\mathcal{W}_p(s, \mathcal{L}_l^i)$. Let $\mathcal{W}_p(s, l) = \cup_{i=1}^{\tau+1} \mathcal{W}_p(s, \mathcal{L}_l^i)$. The size of $\mathcal{W}_p(s, \mathcal{L}_l^i)$ is $|\mathcal{W}_p(s, \mathcal{L}_l^i)| = \tau + 1$ and the number of selected substrings is $|\mathcal{W}_p(s, l)| = (\tau + 1)^2$. The position-aware method satisfies completeness as formalized in Theorem 1.

THEOREM 1. *The position-aware substring selection method satisfies completeness.*

## 4.2 Multi-match-aware Substring Selection

We have an observation that string $s$ may have multiple substrings that match some segments of string $r$. In this case we can discard some of these substrings. For example, consider $r =$ "`vankatesh`" with four segments, {`va`, `nk`, `at`, `esh`}. $s =$ "`avataresha`" has three substrings `va`, `at`, `esh` matching the segments of $r$. We can discard some of these substrings. To this end, we propose a multi-match-aware substring selection method.

Consider $\mathcal{L}_l^i$. Suppose string $s$ has a substring $s_m$ that matches a segment in $\mathcal{L}_l^i$. If we know that $s$ *must have* a substring after $s_m$ which will match a segment in $\mathcal{L}_l^j (j > i)$, we can discard substring $s_m$. For example, $s =$ "`avataresha`" has a substring "`va`" matching a segment in $r =$ "`vankatesh`". Consider the three parts $r_m = s_m =$ "`va`", $r_l = \phi$ and $s_l =$ "`a`", and $r_r =$ "`nkatesh`" and $s_r =$ "`taresha`". As $d_l \geq 1$, if $s$ and $r$ are similar, $d_r \leq \tau - d_l \leq \tau - 1 = 2$. As there are still 3 segments in $r_r$, thus $s_r$ must have a substring matching a segment in $r_r$ based on the pigeon-hole principle. Thus we can discard the substring "va" and use the next substring to find similar pairs. Next we generalize our idea.

Suppose $s$ has a substring $s_m$ with start position $p$ matching a segment $r_m \in \mathcal{L}_l^i$. We still consider the three parts of the two strings: $s_l, s_m, s_r$ and $r_l, r_m, r_r$ as illustrated in Figure 5. Let $\triangle_l = |p_i - p|$. $d_l = \text{ED}(r_l, s_l) \geq \triangle_l$. As there are $i-1$ segments in $s_l$, if each segment only has 1 error when transforming $r_l$ to $s_l$, we have $\triangle_l \leq i-1$. If $\triangle_l \geq i$, $d_l = \text{ED}(r_l, s_l) \geq \triangle_l \geq i$, $d_r = \text{ED}(r_r, s_r) \leq \tau - d_l \leq \tau - i$ (if $s$ is similar to $r$). As $r_r$ contains $\tau + 1 - i$ segments, $s_r$ must contain a substring matching a segment in $r_r$ based on the pigeon-hole principle, which can be proved similar to Lemma 1. In this way, we can discard $s_m$, since for **any** string $r \in \mathcal{L}_l^i(r_m)$, $s$ must have a substring that matches a segment in the right part $r_r$, and thus we can identify strings similar to $s$ using the next matching segment. In summary, if $\triangle_l = |p - p_i| \leq i-1$, we keep the substring with start position $p$ for $\mathcal{L}_l^i$. That is the minimal start position is $\bot_i^l = \max(1, p_i - (i-1))$ and the maximal start position is $\top_i^l = \min(|s| - l_i + 1, p_i + (i-1))$.

For example, consider string $r$="vankatesh" with four segments, {va, nk, at, esh}, and string $s$="avataresha". For the first segment, we have $\bot_i^l$=1-0=1 and $\top_i^l$=1+0=1. Thus the selected substring is only "av" for the first segment. For the second segment, we have $\bot_i^l$=3-1=2 and $\top_i^l$=3+1=4. Thus the selected substrings are "va", "at", and "ta" for the second segment. Similarly for the third segment, we have $\bot_i^l$=5-2=3 and $\top_i^l$=5+2=7, and for the fourth segment, we have $\bot_i^l$=7-3=4 and $\top_i^l$=7+3=10.

The above observation is made from the left-side perspective. Similarly, we can use the same idea from the right-side perspective. As there are $\tau + 1 - i$ segments on the right part $r_r$, there are at most $\tau + 1 - i$ edit errors on $r_r$. If we transform $r$ to $s$ from the right-side perspective, position $p_i$ on $r$ should be aligned with position $p_i + \triangle$ on $s$ as shown in Figure 5(b). Suppose the position $p$ on $s$ matching position $p_i$ on $r$. Let $\triangle_r = |p - (p_i + \triangle)|$. We have $d_r = \text{ED}(s_r, r_r) \geq \triangle_r$. As there are $\tau + 1 - i$ segments on the right part $r_r$, we have $\triangle_r \leq \tau + 1 - i$. Thus the minimal start position for $\mathcal{L}_l^i$ is $\bot_i^r = \max(1, p_i + \triangle - (\tau + 1 - i))$ and the maximal start position is $\top_i^r = \min(|s| - l_i + 1, p_i + \triangle + (\tau + 1 - i))$.

Consider the above example. Suppose $\tau = 3$ and $\triangle = 1$. For the fourth segment, we have $\bot_i^r = 7+1-(3+1-4) = 8$ and $\top_i^r = 7+1+(3+1-4) = 8$. Thus the selected substring is only "sha" for the fourth segment. Similarly for the third segment, we have $\bot_i^r = 5$ and $\top_i^r = 7$. Thus the selected substrings are "ar", "re", and "es" for the third segment.

More interestingly, we can use the two techniques simultaneously. That is for $\mathcal{L}_l^i$, we only select the substrings with the start positions between $\bot_i = \max(\bot_i^l, \bot_i^r)$ and $\top_i = \min(\top_i^l, \top_i^r)$ and with length $l_i$, denoted by $\mathcal{W}_m(s, \mathcal{L}_l^i)$. Let $\mathcal{W}_m(s, l) = \cup_{i=1}^{\tau+1} \mathcal{W}_m(s, \mathcal{L}_l^i)$. The number of selected substrings is $|\mathcal{W}_m(s, l)| = \lfloor \frac{\tau^2 - \triangle^2}{2} \rfloor + \tau + 1$ as stated in Lemma 2.

LEMMA 2. $|\mathcal{W}_m(s, l)| = \lfloor \frac{\tau^2 - \triangle^2}{2} \rfloor + \tau + 1$.

Moreover we prove that the multi-match-aware selection method satisfies completeness as stated in Theorem 2.

THEOREM 2. *The multi-match-aware substring selection method satisfies completeness.*

Consider the above example. For the first segment, we have $\bot_i = 1 - 0 = 1$ and $\top_i = 1 + 0 = 1$. We select "av" for the first segment. For the second segment, we have $\bot_i = 3 - 1 = 2$ and $\top_i = 3 + 1 = 4$. We select substrings "va", "at", and "ta" for the second segment. For the third segment, we have $\bot_i = 5 + 1 - (3 + 1 - 3) = 5$ and $\top_i = 5 + 1 + (3 + 1 - 3) = 7$. We select substrings "ar", "re", and "es" for the third segment. For the fourth segment, we have $\bot_i = 7 + 1 - (3 + 1 - 4) = 8$ and $\top_i = 7 + 1 + (3 + 1 - 4) = 8$. Thus we select the substring "sha" for the fourth segment. The multi-match-aware method only selects 8 substrings.

### 4.3 Comparison of Selection Methods

We compare the selected substring sets of different methods. Let $\mathcal{W}_\ell(s,l), \mathcal{W}_f(s,l), \mathcal{W}_p(s,l), \mathcal{W}_m(s,l)$ respectively denote the sets of selected substrings that use the length-based selection method, the shift-based selection method, the position-aware selection method, and the multi-match-aware selection method. Based on the size analysis of each set, we have $|\mathcal{W}_m(s,l)| \leq |\mathcal{W}_p(s,l)| \leq |\mathcal{W}_f(s,l)| \leq |\mathcal{W}_\ell(s,l)|$. Next we prove $\mathcal{W}_m(s,l) \subseteq \mathcal{W}_p(s,l) \subseteq \mathcal{W}_f(s,l) \subseteq \mathcal{W}_\ell(s,l)$ as formalized in Lemma 3.

LEMMA 3. *For any string $s$ and a length $l$, we have $\mathcal{W}_m(s,l) \subseteq \mathcal{W}_p(s,l) \subseteq \mathcal{W}_f(s,l) \subseteq \mathcal{W}_\ell(s,l)$.*

Moreover, we can prove that $\mathcal{W}_m(s,l)$ has the minimum size among all substring sets generated by the methods that satisfy completeness as formalized in Theorem 3.

THEOREM 3. *The substring set $\mathcal{W}_m(s,l)$ generated by the multi-match-aware selection method has the minimum size among all the substring sets generated by the substring selection methods that satisfy completeness.*

Theorem 3 proves that the substring set $\mathcal{W}_m(s,l)$ has the minimum size. Next we introduce another concept to show the superiority of our multi-match-aware selection method.

DEFINITION 3 (MINIMALITY). *A substring set $\mathcal{W}(s,l)$ generated by a method with the completeness property satisfies minimality, if for any substring set $\mathcal{W}'(s,l)$ generated by a method with the completeness property, $\mathcal{W}(s,l) \subseteq \mathcal{W}'(s,l)$.*

Next we prove that if $l \geq 2(\tau+1)$ and $|s| \geq l$, the substring set $\mathcal{W}_m(s,l)$ generated by our multi-match-aware selection method satisfies minimality as stated in Theorem 4. The condition $l \geq 2(\tau + 1)$ makes sense where each segment is needed to have at least two characters. For example, if $10 \leq l < 12$, we can tolerate $\tau = 4$ edit operations. If $12 \leq l < 14$, we can tolerate $\tau = 5$ edit operations.

THEOREM 4. *If $l \geq 2(\tau+1)$ and $|s| \geq l$, $\mathcal{W}_m(s,l)$ satisfies minimality.*

### 4.4 Substring-selection Algorithm

Based on above discussion, we improve SUBSTRINGSELECTION algorithm by avoiding unnecessary substrings. For $\mathcal{L}_l^i$, we use the multi-match-aware selection method to select substrings, and the selection complexity is $\mathcal{O}(\tau)$. Figure 6 gives the pseudo-code of the selection algorithm.

For example, consider the strings in Table 1. We create inverted indices as illustrated in Figure 1. Consider string $s_1$ = "vankatesh" with four segments, we build four inverted lists for its segments {va, nk, at, esh}. Then for $s_2$ = "avataresha". We use multi-match-aware selection method to select its substrings. Here we only select 8 substrings for $s_2$ and use the 8 substrings to find similar strings of $s_2$ from the inverted indices. Similarly, we can select substrings and find similar string pairs for other strings.

**Algorithm 2**: SUBSTRINGSELECTION($s, \mathcal{L}_l^i$)

**Input**: $s$: A string; $\mathcal{L}_l^i$: Inverted index
**Output**: $\mathcal{W}(s, \mathcal{L}_l^i)$: Selected substrings
1 **begin**
2    **for** $p \in [\bot_i, \top_i]$ **do**
3          Add the substring of $s$ with start position $p$ and with length $l_i$ ($s[p, l_i]$) into $\mathcal{W}(s, \mathcal{L}_l^i)$;
4 **end**

**Figure 6: SubstringSelection algorithm**

## 5. IMPROVING THE VERIFICATION

In our framework, for string $s$ and inverted index $\mathcal{L}_l^i$, we generate a set of its substrings $\mathcal{W}(s, \mathcal{L}_l^i)$. For each substring $w \in \mathcal{W}(s, \mathcal{L}_l^i)$, we need to check whether it appears in $\mathcal{L}_l^i$. If $w \in \mathcal{L}_l^i$, for each string $r \in \mathcal{L}_l^i(w)$, $\langle r, s \rangle$ is a candidate pair and we need to verify the candidate pair. In this section we propose effective techniques to do efficient verification.

### 5.1 Length-aware Verification

Given a candidate pair $\langle r, s \rangle$, a straightforward method to verify the pair is to use a dynamic-programming algorithm to compute their real edit distance. If the edit distance is not larger than $\tau$, the pair is an answer. We can use a matrix $M$ with $|r| + 1$ rows and $|s| + 1$ columns to compute their edit distance, in which $M(0, j) = j$ for $0 \leq j \leq |s|$, and for $1 \leq i \leq |r|$ and $0 \leq j \leq |s|$,

$$M(i,j) = \min\bigl(M(i-1,j)+1, M(i,j-1)+1, M(i-1,j-1)+\delta\bigr)$$

where $\delta = 0$ if the $i$-th character of $r$ is the same as the $j$-th character of $s$; otherwise $\delta = 1$. The time complexity of the dynamic-programming algorithm is $\mathcal{O}(|r| * |s|)$.

Actually, we do not need to compute their real edit distance and only need to check whether their edit distance is not larger than $\tau$. An improvement based on length pruning [20] is proposed which only computes the values $M(i, j)$ for $|i - j| \leq \tau$, as shown in the shaded cells of Figure 7(a). The basic idea is that if $|i - j| > \tau$, $M(i, j) > \tau$, and we do not need to compute such values. This method improves the time complexity $\mathcal{V}(s, r)$ to $\mathcal{O}((2*\tau+1)*\min(|r|, |s|))$. Next, we propose a technique to further improve the performance by considering the length difference between $r$ and $s$.

We first use an example to illustrate our idea. Consider string $r =$ "kaushuk chadhui" and string $s =$ "caushik chakrabar". Suppose $\tau = 3$. Existing methods need to compute all the shaded values in Figure 7(a). We have an observation that we do not need to compute $M(2, 1)$, which is the edit distance between "ka" and "c". This is because if there is a transformation from $r$ to $s$ by first transforming "ka" to "c" with at least 1 edit operation (length difference) and then transforming "ushuk chadhui" to "aushik chakrabar" with at least 3 edit operations (length difference), the transformation distance is at least 4 which is larger than $\tau = 3$. In other words, even if we do not compute $M(2, 1)$, we know that there is no transformation including $M(2, 1)$ (the transformation from "ka" to "c") whose edit distance is not larger than $\tau$. Actually we only need to compute the highlighted values as illustrated in Figure 7(b).

To address this problem, we propose a length-aware verification method. Without loss of generality, let $|s| \geq |r|$ and $\triangle = |s| - |r| \leq \tau$ (otherwise their edit distance must be larger than $\tau$). We call a transformation from $r$ to $s$ including $M(i, j)$, if the transformation first transforms the first $i$ characters of $r$ to the first $j$ characters of $s$ with $d_1$ edit operations and then transforming the other characters in $r$ to the other characters in $s$ with $d_2$ edit operations. Based on length difference, we have $d_1 \geq |i - j|$ and $d_2 \geq |(|s| - j) - (|r| - i)| = |\triangle + (i - j)|$. If $d_1 + d_2 > \tau$, we do not need to compute $M(i, j)$, since the distance of any transformation including $M(i, j)$ is larger than $\tau$. To check whether $d_1 + d_2 > \tau$, we consider the following cases.

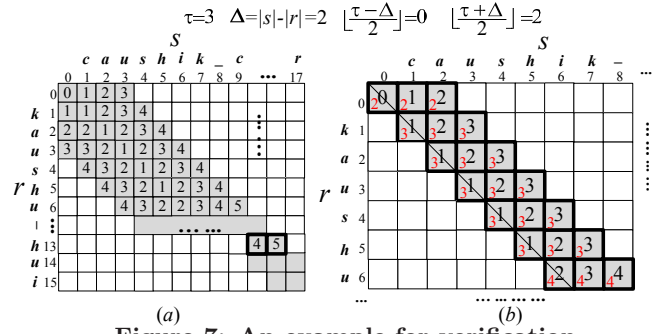

**Figure 7: An example for verification**

(1) If $i \geq j$, we have $d_1 + d_2 \geq i - j + \triangle + i - j$. If $i - j + \triangle + i - j > \tau$, that is $j < i - \frac{\tau - \triangle}{2}$, we do not need to compute $M(i, j)$. In other words, we only need to compute $M(i, j)$ with $j \geq i - \frac{\tau - \triangle}{2}$.

(2) If $i < j$, $d_1 = j - i$. If $j - i \leq \triangle$, $d_1 + d_2 \geq j - i + \triangle - (j - i) = \triangle$. As $\triangle \leq \tau$, there is no position constraint. We need to compute $M(i, j)$; otherwise if $j - i > \triangle$, we have $d_1 + d_2 \geq j - i + j - i - \triangle$. If $j - i + j - i - \triangle > \tau$, that is $j > i + \frac{\tau + \triangle}{2}$, we do not need to compute $M(i, j)$. In other words, we only need to compute $M(i, j)$ with $j \leq i + \frac{\tau + \triangle}{2}$.

Based on this observation, for each row $M(i, *)$, we only compute $M(i, j)$ for $i - \lfloor \frac{\tau - \triangle}{2} \rfloor \leq j \leq i + \lfloor \frac{\tau + \triangle}{2} \rfloor$. For example, in Figure 8, we only need to compute the values in black circles. Thus we can improve the time complexity $\mathcal{V}(s, r)$ from $\mathcal{O}((2\tau+1)*\min(|r|, |s|))$ to $\mathcal{O}((\tau+1)*\min(|r|, |s|))$.

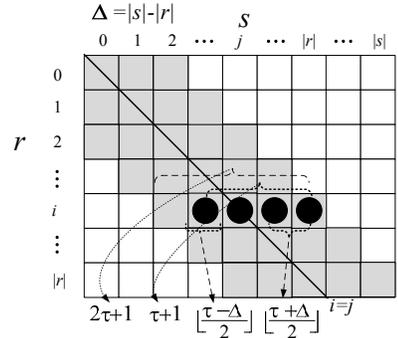

**Figure 8: Length-aware verification**

**Early Termination:** We can further improve the performance by doing an early termination. Consider the values in row $M(i, *)$. A straightforward early-termination method is to check each value in $M(i, *)$, and if each value is larger than $\tau$, we can do an early termination. This is because the values in the following rows $M(k > i, *)$ must be larger than $\tau$ based on the dynamic-programming algorithm. This pruning technique is called *prefix pruning*. For example in Figure 7(a), if $\tau = 3$, after we have computed $M(13, *)$, we can do an early termination as all the values in $M(13, *)$ are larger than $\tau$. But in our method, after we have computed the values in $M(6, *)$, we can conclude that the edit distance between the two strings is at least 4 (larger than $\tau = 3$). Thus we do not need to compute $M(i > 6, *)$ and can terminate the computation as shown in Figure 7(b). To this end, we propose a novel early-termination method.

For ease of presentation, we first introduce several notations. Given a string $s$, let $s[i]$ denote the $i$-th character and $s[i:j]$ denote the substring of $s$ from the $i$-th character to the $j$-th character. Notice that $M(i,j)$ denotes the edit distance between $r[1:i]$ and $s[1:j]$. We can estimate the lower bound of the edit distance between $r[i:|r|]$ and $s[j:|s|]$ using their length difference $\big|(|s|-j)-(|r|-i)\big|$. We use $E(i,j) = M(i,j) + \big|(|s|-j) - (|r|-i)\big|$ to estimate the edit distance between $s$ and $r$, which is called *expected edit distance* of $s$ and $r$ with respect to $M(i,j)$. If each expected edit distance for $M(i,j)$ in $M(i,*)$ is larger than $\tau$, the edit distance between $r$ and $s$ must be larger than $\tau$, thus we can do an early termination. To achieve our goal, for each value $M(i,j)$, we maintain the expected minimal edit distance $E(i,j)$. If each value in $E(i,*)$ is larger than $\tau$, we can do an early termination as formalized in Lemma 4.

LEMMA 4. *Given strings $s$ and $r$, if each value in $E(i,*)$ is larger than $\tau$, the edit distance of $r$ and $s$ is larger than $\tau$.*

For example, in Figure 7(b), we show the expected edit distances in the left-bottom corner of each cell. When we have computed $M(6,*)$ and $E(6,*)$, all values in $E(6,*)$ are larger than 3, thus we can do an early termination. In this way, we can avoid many unnecessary computations. Note that our proposed verification techniques can be applied to any other algorithms to verify a candidate pair in terms of edit distance (e.g., ED-JOIN and NGPP).

## 5.2 Extension-based Verification

Consider a selected substring $w$ of string $s$. If $w$ appears in the inverted index $\mathcal{L}_l^i$, for each string $r$ in the inverted list $\mathcal{L}_l^i(w)$, we need to verify the pair $\langle s, r \rangle$. As $s$ and $r$ share a common segment $w$, we can use the shared segment to efficiently verify the pair. To achieve our goal, we propose an extension-based verification algorithm.

As $r$ and $s$ share a common segment $w$, we partition them into three parts based on the common segment. We partition $r$ into three parts, the left part $r_l$, the matching part $r_m = w$, and the right part $r_r$. Similarly, we get three parts for string $s$: $s_l$, $s_m = w$, and $s_r$. Here we align $s$ and $r$ based on the matching substring $r_m$ and $s_m$, and we only need to verify whether $r$ and $s$ are similar in this alignment. Thus we first compute the edit distance $d_l = \text{ED}(r_l, s_l)$ between $r_l$ and $s_l$ using the above-mentioned method. If $d_l$ is larger than $\tau$, we terminate the computation; otherwise, we compute the edit distance $d_r = \text{ED}(s_r, r_r)$ between $s_r$ and $r_r$. If $d_l + d_r$ is larger than $\tau$, we discard the pair; otherwise we take it as an answer. Note that this method will not involve false negatives. This is because based on Lemma 1, if $s$ and $r$ are similar, $s$ must have a substring that matches a segment of $r$. In addition, based on dynamic-programming algorithm, there must exist a transformation by aligning $r_m$ with $s_m$ and $\text{ED}(s, r) = d_l + d_r$. As our method selects all possible substrings and considers all such common segments, our method will not miss any results. On the other hand, as we find the answers with $d_l + d_r \leq \tau$ and $\text{ED}(s, r) \leq d_l + d_r \leq \tau$, our method will not involve false positives. To guarantee correctness of our extension-based method, we first give a formal definition of correctness.

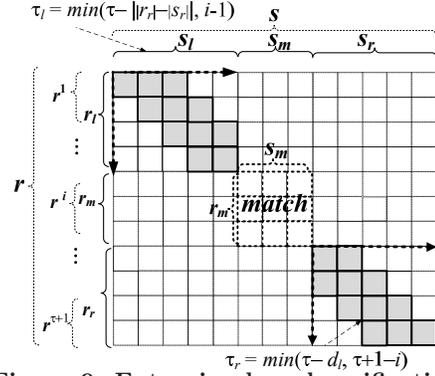

Figure 9: Extension-based verification

DEFINITION 4 (CORRECTNESS). *Given a candidate pair $\langle s, r \rangle$, a verification algorithm is correct, if it satisfies (1) If $\langle s, r \rangle$ passes the algorithm, $\langle s, r \rangle$ must be a similar pair; and (2) If $\langle s, r \rangle$ is a similar pair, it must pass the algorithm.*

We prove that our extension-based verification method satisfies correctness as stated in Theorem 5.

THEOREM 5. *Our extension-based verification method satisfies correctness.*

Actually, we can further improve the verification algorithm. For the left parts, we can give a tighter threshold $\tau_l \leq \tau$. The basic idea is as follows. As the minimal edit distance between the two right segments $r_r$ and $s_r$ is $\big||r_r|-|s_r|\big|$. Thus we can set $\tau_l = \tau - \big||r_r| - |s_r|\big|$. If the edit distance between $r_l$ and $s_l$ is larger than threshold $\tau_l$, we can terminate the verification; otherwise we continue to compute $d_r = \text{ED}(r_r, s_r)$. Similarly for the right parts, we can also give a tighter threshold $\tau_r \leq \tau$. As $d_l$ has been computed, we can set $\tau_r = \tau - d_l$ as a threshold to verify whether $r_r$ and $s_r$ are similar. If $d_r$ is larger than threshold $\tau_r$, we can terminate the verification.

For example, if we want to verify $s_5 = $ "`kaushuk chadhui`" and $s_6 = $ "`caushik chakrabar`". $s_5$ and $s_6$ share a segment "`_cha`". We have $s_{5_l} = $ "`kaushuk`" and $s_{6_l} = $ "`caushik`", and $s_{5_r} = $ "`dhui`" and $s_{6_r} = $ "`krabar`". Suppose $\tau = 3$. As $\big||s_{5_r}| - |s_{6_r}|\big| = 2$, $\tau_l = \tau - 2 = 1$. We only need to verify whether the edit distance between $s_{5_l}$ and $s_{6_l}$ is not larger than $\tau_l = 1$. After we have computed $M(6,*)$, we can do an early termination as each value in $E(6,*)$ is larger than 1, as shown in Figure 7. Note that as $\tau_l = 1$ and $|s_{5_l}| - |s_{6_l}| = 0$, $\bot_i = \top_i = 0$. Thus we only need to compute $M(i,i)$.

We discuss how to deduce a tighter bound for $\tau_l$ and $\tau_r$. Consider the $i$-th segment. If $d_l \geq i$, we can terminate the verification based on the multi-match-aware method. Thus we have $\tau_l = i - 1$. Combining with the above pruning condition, we have $\tau_l = \min(\tau - \big||r_r| - |s_r|\big|, i-1)$. As $\big||r_r| - |s_r|\big| = \big|(|r| - p_i - l_i) - (|s| - p - l_i)\big| = |p - p_i - \triangle| \leq \tau + 1 - i$ (based on the multi-match-aware method), $\tau - \big||r_r| - |s_r|\big| \geq i - 1$. We set $\tau_l = i - 1$. Similarly we have $\tau_r = \min(\tau - d_l, \tau + 1 - i)$. As $d_l \leq \tau_l \leq i - 1$, $\tau - d_l \geq \tau - (i-1)$. Thus we set $\tau_r = \tau + 1 - i$.

## 5.3 Sharing Computations

Given a selected substring $w$, there may be large numbers of strings in $\mathcal{L}_l^i(w)$ similar to string $s$. When computing the edit distance of the left parts $s_l$ and $r_l$ (and that of the right parts $s_r$ and $r_r$), we can share the computations if they have common prefixes. Next we discuss how to share computations. As the strings in $\mathcal{L}_l^i(w)$ are sorted in alphabetical

**Algorithm 3**: VERIFICATION($s, \mathcal{L}_l^i(w), \tau$)

**Input**: $s$: A string; $\mathcal{L}_l^i(w)$: Inverted list; $\tau$: Threshold
**Output**: $\mathcal{R} = \{(s \in \mathcal{S}, r \in \mathcal{S}) \mid \text{ED}(s, r) \leq \tau\}$
1 **begin**
2  $\quad \tau_l = i - 1$;
3  $\quad \tau_r = \tau + 1 - i$;
4  $\quad$ **for** $r \in \mathcal{L}_l^i(w)$ **do**
5  $\quad\quad d_l = $ VERIFYSTRINGPAIR($s_l, r_l, \tau_l$);
6  $\quad\quad$ **if** $d_l \leq \tau_l$ **then**
7  $\quad\quad\quad d_r = $ VERIFYSTRINGPAIR($s_r, r_r, \tau_r$);
8  $\quad\quad\quad$ **if** $d_r \leq \tau_r$ **then** $\mathcal{R} \leftarrow \langle r, s \rangle$;
9 **end**

**Function** VERIFYSTRINGPAIR($s, r, \tau'$)

**Input**: $s$: A string; $r$: A string; $\tau'$: A threshold
**Output**: $d = \min(\tau' + 1, \text{ED}(s, r))$
1 **begin**
2  $\quad$ Using the length-aware verification with the threshold $\tau'$ and sharing the computations on common prefixes;
3  $\quad$ **if** *Early Termination* **then** $d = \tau' + 1$;
4  $\quad$ **else** $d = \text{ED}(s, r)$;
5 **end**

**Figure 10: Verification algorithm**

order, we visit strings in $\mathcal{L}_l^i(w)$ in order. Suppose the first string is $r_1$ and its three parts are $r_{1_l}, r_{1_m}, r_{1_r}$. We compute the edit distance between $r_{1_l}$ and $s_l$ using the dynamic-programming algorithm. We store the matrix for $r_{1_l}$ and $s_l$. For the next string $r_2$ with left part $r_{2_l}$, we use the stored matrix to compute the edit distance between $r_{2_l}$ and $s_l$. We first compute the longest common prefix between $r_{2_l}$ and $r_{1_l}$, denoted by $c$. When computing the edit distance between $s_l$ and $r_{2_l}$, we use the stored matrix on $s_l$ and $c$ which has already been computed for $s_l$ and $r_{1_l}$. Then for the characters after $c$ in $r_{2_l}$, we continue the computation using the kept matrix. Thus we avoid many unnecessary computations. Notice that we do not need to maintain multiple matrixes and only keep a single matrix for the current string. We use the same idea on the right parts($s_r, r_r$).

### 5.4 Verification Algorithm

Based on our proposed techniques, we improve the VERIFICATION algorithm. Consider a string $s$, a selected substring $w$, and an inverted list $\mathcal{L}_l^i(w)$. For $r \in \mathcal{L}_l^i(w)$, we use the extension-based method to verify the candidate pair $\langle s, r \rangle$ as follows. We first compute $\tau_l = i - 1$ and $\tau_r = \tau + 1 - i$. Then for each $r \in \mathcal{L}_l^i(w)$, we compute the edit distance ($d_l$) between $r_l$ and $s_l$ using the tighter bound $\tau_l$. If $d_l > \tau_l$, we terminate the verification; otherwise we verify whether $s_r$ and $r_r$ are similar with threshold $\tau_r$. When computing the edit distance between $s_l$ and $r_l(s_r$ and $r_r)$, we use the length-aware verification and share the computations on common prefixes. Figure 10 illustrates the pseudo-code.

### 5.5 Correctness and Completeness

We prove correctness and completeness of our algorithm as formalized in Theorem 6.

THEOREM 6. *Our algorithm satisfies (1) completeness: Given any similar pair $\langle s, r \rangle$, our algorithm must find it as an answer; and (2) correctness: A pair $\langle s, r \rangle$ found in our algorithm must be a similar pair.*

## 6. EXPERIMENTAL STUDY

We have implemented our method and conducted an extensive set of experimental studies on three real datasets: DBLP Author[†], DBLP Author+Title, and AOL Query Log[‡]. DBLP Author is a dataset with short strings, DBLP Author+Title is a dataset with long strings, and the Query Log is a set of query logs. Table 2 shows the detailed information of the datasets. Note that the Author+Title dataset is the same as that used in ED-JOIN and the Author dataset is the same as that used in TRIE-JOIN. Figure 11 shows the string length distributions of different datasets.

**Table 2: Datasets**

| Datasets | Cardinality | Avg Len | Max Len | Min Len |
|---|---|---|---|---|
| Author | 612781 | 14.826 | 46 | 6 |
| Query Log | 464189 | 44.75 | 522 | 30 |
| Author+Title | 863073 | 105.82 | 886 | 21 |

We compared our algorithms with state-of-the-art methods, ED-JOIN [23] and TRIE-JOIN [20]. As ED-JOIN and TRIE-JOIN outperform other methods, e.g., Part-Enum [2] and All-Pairs-Ed [3] (also experimentally proved in [23, 20]), in the paper we only compared our method with the two best studies. We downloaded their binary codes from their homepages, ED-JOIN [§] and TRIE-JOIN [¶].

All the algorithms were implemented in C++ and compiled using GCC 4.2.4 with -O3 flag. All the experiments were run on a Ubuntu machine with an Intel Core 2 Quad X5450 3.00GHz processor and 4 GB memory.

### 6.1 Evaluating Substring Selection

In this section, we evaluate substring selection techniques. We implemented the following four methods. (1) The length-based selection method, denoted by Length, which selects the substrings with the same lengths as the segments. (2) The shift-based method, denoted by Shift, which selects the substring by shifting $[-\tau, \tau]$ positions as discussed in Section 4. (3) Our position-aware selection method, denoted by Position. (4) Our multi-match-aware selection method, denoted by Multi-match. We first compared the total number of selected substrings. Figure 12 shows the results.

We can see that the Length-based method selected large numbers of substrings. The number of selected substring of the Position-based method was about a tenth to a fourth of that of the Length-based method and a half of the Shift-based method. The Multi-match-based method further reduced the number of selected substrings to about a half of that of the Position-based method. For example, on Author dataset, for $\tau = 1$, the Length-based method selected 19 million substrings, the Shift-based method selected 5.5 million substrings, the Position-based method reduced the number to 3.7 million, and the Multi-match-based method further deceased the number to 2.4 million. Based on our analysis in Section 4, for strings with $l$, the length-based method selected $(\tau+1)(|s|+1)-l$ substrings, the shift-based method selected $(\tau + 1)(2\tau + 1)$ substrings, the position-based method selected $(\tau + 1)^2$ substrings, and the multi-match-aware method selected $\lfloor \frac{\tau^2 - \triangle^2}{2} \rfloor + \tau + 1$ substrings. If $|s|=l=15$ and $\tau = 1$, the number of selected substrings of the four methods are respectively 17, 6, 4, and 2. Obviously the experimental results consisted with our theoretical analysis.

[†]http://www.informatik.uni-trier.de/~ley/db
[‡]http://www.gregsadetsky.com/aol-data/
[§]http://www.cse.unsw.edu.au/~weiw/project/simjoin.html
[¶]http://dbgroup.cs.tsinghua.edu.cn/wangjn/

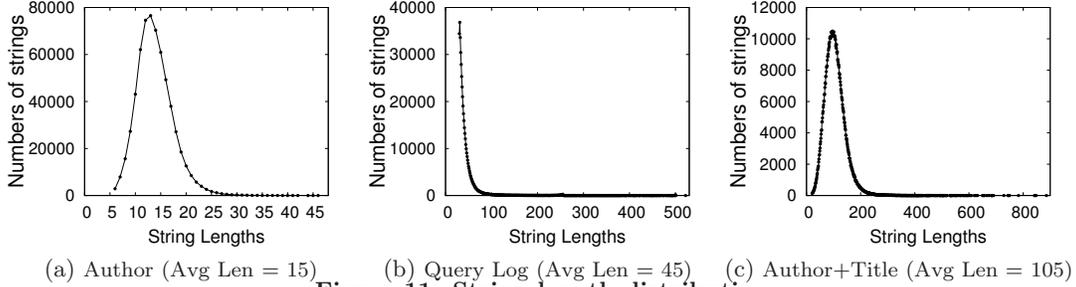

(a) Author (Avg Len = 15)  (b) Query Log (Avg Len = 45)  (c) Author+Title (Avg Len = 105)

Figure 11: String length distribution

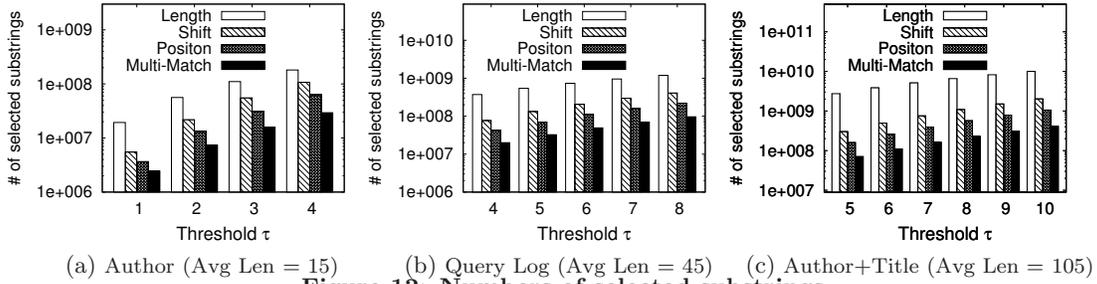

(a) Author (Avg Len = 15)  (b) Query Log (Avg Len = 45)  (c) Author+Title (Avg Len = 105)

Figure 12: Numbers of selected substrings

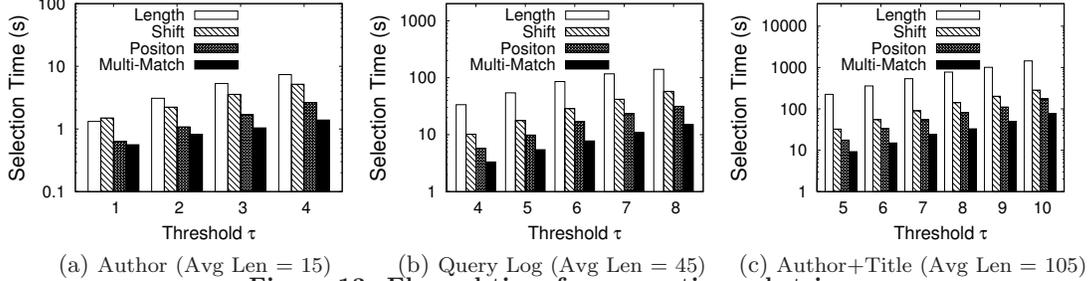

(a) Author (Avg Len = 15)  (b) Query Log (Avg Len = 45)  (c) Author+Title (Avg Len = 105)

Figure 13: Elapsed time for generating substrings

We also compared the elapsed time to generate substrings. Figure 13 shows the results. We see that the Multi-match-based method outperformed the Position-based method which in turns was better than the Shift-based method and the Length-based method. This is because the elapsed time depended on the number of selected substrings and the Multi-match-based selected the smallest number of substrings.

## 6.2 Evaluating Verification

In this section, we evaluate our verification techniques. We implemented four methods. (1) The naive method, denoted by $2\tau+1$, which computed $2\tau+1$ values in each row and used the naive early-termination technique (if all values in a row are larger than $\tau$, we terminate). (2) Our length-aware method, denoted by $\tau+1$, which computed $\tau+1$ values in each row and used the expected edit distance to do early termination. (3) Our extension-based method, denoted by Extension, which used the extension-based framework. It also computed $\tau+1$ rows and used the expected edit distance to do early termination. (4) We used the extension-based method with sharing computations on common prefixes, denoted by SharePrefix. Figure 14 shows the results.

We see that the naive method had the worst performance, as it needed to compute many unnecessary values in the matrix. Our length-aware method was 2-5 times faster than the naive method. This is because our length-aware method can decrease the complexity from $2\tau+1$ to $\tau+1$ and used expected edit distances to do early termination. The extension-based method achieved higher performance and was 2-4 times faster than the length-aware method. The reason is that the extension-based method can avoid the duplicated computations on the common segments and it also used a tighter bound to verify the left parts and the right parts. The SharePrefix method achieved the best performance, as it can avoid many unnecessary computations for strings with common prefixes. For example, on the Author dataset, for $\tau = 4$ the naive method took 10,000 seconds, the length-aware method decreased the time to 4000 seconds, the extension-based method reduced it to 2000 seconds, and the SharePrefix method further improved the time to about 700 seconds. On the Query Log dataset, for $\tau = 8$, the elapsed time of the four methods were respectively 3500 seconds, 1500 seconds, 600 seconds, and 450 seconds.

## 6.3 Comparison with Existing Methods

In this section, we compare our method with state-of-the-art methods ED-JOIN [23] and TRIE-JOIN [20]. As TRIE-JOIN had multiple algorithms, we reported the best results. For ED-JOIN, we tuned its parameter $q$ and reported the best results. As TRIE-JOIN was efficient for short strings, we downloaded the same dataset from TRIE-JOIN homepage (i.e., Author with short strings) and used it to compare with TRIE-JOIN. As ED-JOIN was efficient for long strings, we downloaded the same dataset from ED-JOIN homepage (i.e., Author+Title with long strings) and used it to compare with ED-JOIN. Figure 15 shows the results, where the elapsed time included the indexing time and the join time.

On the Author dataset with short strings, TRIE-JOIN outperformed ED-JOIN, and our method was much better than

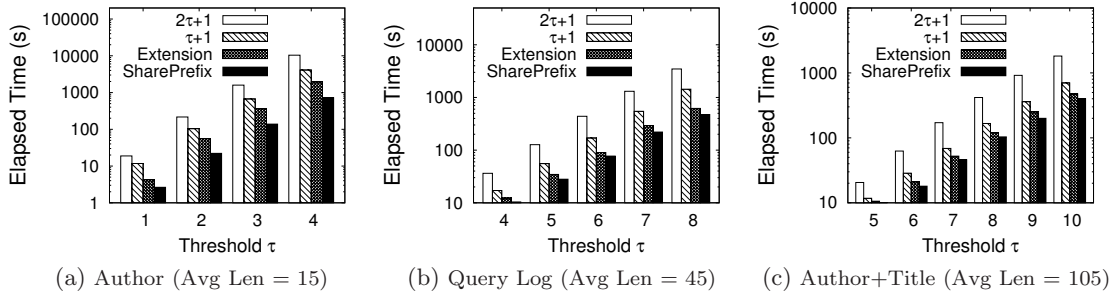
(a) Author (Avg Len = 15)  (b) Query Log (Avg Len = 45)  (c) Author+Title (Avg Len = 105)
Figure 14: Elapsed time for verification

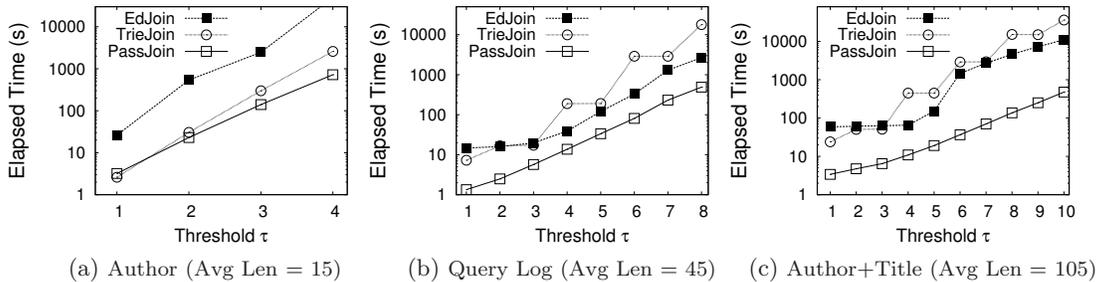
(a) Author (Avg Len = 15)  (b) Query Log (Avg Len = 45)  (c) Author+Title (Avg Len = 105)
Figure 15: Comparison with state-of-the-art methods

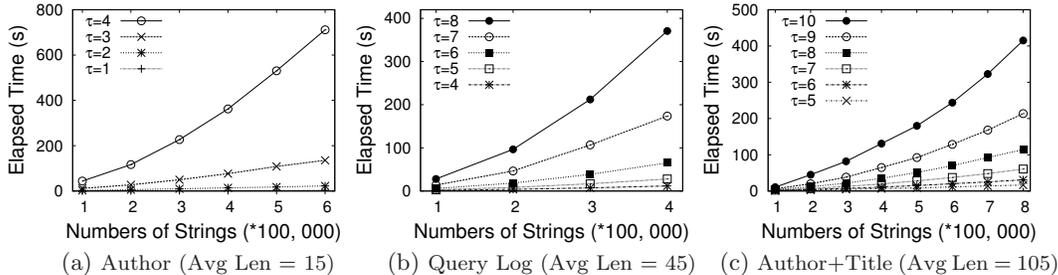
(a) Author (Avg Len = 15)  (b) Query Log (Avg Len = 45)  (c) Author+Title (Avg Len = 105)
Figure 16: Scalability

them, especially for $\tau \geq 2$. The main reason is as follows. ED-JOIN must use a smaller $q$ for a larger threshold. In this way ED-JOIN will involve large numbers of candidate pairs, since a smaller $q$ has rather lower pruning power [23]. TRIE-JOIN used the prefix filtering to find similar pairs using a trie structure. If a small number of strings shared prefixes, TRIE-JOIN had low pruning power and was expensive to traverse the trie structure. Instead our framework utilized segments to prune large numbers of dissimilar pairs. The segments were selected across the strings and not restricted to prefix filtering. For instance, for $\tau=4$, TRIE-JOIN took 2500 seconds. PASS-JOIN improved it to 700 seconds. ED-JOIN was rather slow and even larger than 10,000 seconds.

On the Author+Title dataset with long strings, our method significantly outperformed ED-JOIN and TRIE-JOIN, even in 2-3 orders of magnitude. This is because TRIE-JOIN was rather expensive to traverse the trie structures with long strings, especially for large thresholds. ED-JOIN needed to use a mismatch technique to do pruning which was inefficient while our filtering algorithm is very efficient. In addition, our verification method was more efficient than existing ones. For instance, for $\tau = 8$, TRIE-JOIN took 15,000 seconds, ED-JOIN decreased it to 5000 seconds, and PASS-JOIN improved the time to 130 seconds.

In addition, we compared index sizes on three datasets, as shown in Table 3. We can observe that existing methods involve larger indices than our method. For example, on the Author+Title dataset, ED-JOIN had 335 MB index,

TRIE-JOIN used 90 MB, and our method only took 2.1 MB. There are two main reasons. Firstly for each string with length $l$, ED-JOIN generated $l-q+1$ grams where $q$ is the gram length, and our method only generated $\tau+1$ segments. Secondly for a string with length $l$, we only maintained the indices for strings with lengths between $l-\tau$ and $l$, and ED-JOIN kept indices for all strings. TRIE-JOIN needed to use a trie structure to maintain strings, which had overhead to store the strings (e.g., pointers to children and indices for searching children with a given character).

Table 3: Index sizes (MB)

| Data Sets | Data Sizes | ED-JOIN ($q = 4$) | TRIE-JOIN | PASS-JOIN ($\tau = 4$) |
|---|---|---|---|---|
| Author | 8.7 | 25.34 | 16.32 | 1.92 |
| Query Log | 20 | 72.17 | 69.65 | 4.96 |
| Author+Title | 88 | 335.24 | 90.17 | 2.1 |

### 6.4 Scalability

In this section, we tested the scalability of our method. We varied the number of strings in the dataset and tested the elapsed time. Figure 16 shows the results. We can see that our method achieved nearly linear scalability. For example, for $\tau = 4$, on the Author dataset, the elapsed time for 400,000 strings, 500,000 strings, and 600,000 strings were respectively 360 seconds, 530 seconds, and 700 seconds.

### 7. RELATED WORK

There have been many studies on string similarity joins [7, 2, 3, 6, 18, 23, 24, 19]. The approaches most related to ours are TRIE-JOIN [20], All-Pairs-Ed [3], ED-JOIN [23], and Part-Enum [2]. All-Pairs-Ed is a $q$-gram-based method. It first generates $q$-grams for each string and then selects the first $q\tau + 1$ grams as a gram prefix based on a pre-defined order. It prunes the string pairs with no common grams and verifies the survived string pairs. ED-JOIN improves All-Pairs-Ed using location-based and content-based mismatch filter by decreasing the number of grams. It has been shown that ED-JOIN outperforms All-Pairs-Ed [3]. TRIE-JOIN uses a trie structure to do similarity joins using prefix filtering. Part-Enum proposed an effective signature scheme called Part-Enum to do similar joins for hamming distance. It has been proved that All-Pairs-Ed and Part-Enum are worse than ED-JOIN and TRIE-JOIN [20]. Thus we only compared with ED-JOIN and TRIE-JOIN.

Gravano et al. [7] proposed gram-based methods and used SQL statements for similarity joins inside relational databases. Sarawagi et al. [18] proposed inverted index-based algorithms to solve similarity-join problem. Chaudhuri et al. [6] proposed a primitive operator for effective similarity joins. Arasu et al. [2] developed a signature scheme which can be used as a filter for effective similarity joins. Xiao et al. [25] proposed ppjoin to improve all-pair algorithm by introducing positional filtering and suffix filtering. Xiao et al. [24] studied top-$k$ similarity joins, which can directly find the top-$k$ string pairs without a given threshold.

In addition, Jacox et al. [11] studied the metric-space similarity join. As this method is not as efficient as ED-JOIN and TRIE-JOIN [20], we did not compare with it in the paper. Chaudhuri et al. [6] proposed the prefix-filtering signature scheme for effective similarity join. Recently, Wang et al. [21] devised a new similarity function by tolerating token errors in token-based similarity and developed effective algorithms to support similarity join on such functions.

The other related studies are approximate string searching [5, 14, 8, 9, 26], which given a query string and a set of strings, finds all similar strings of the query string in the string set. Navarro studied the approximate string matching problem [17], which given a query string and a text string, finds all substrings of the text string that are similar to the query string. These two problems are different from our similarity-join problem, which given two sets of strings, finds all similar string pairs. There are some studies on selectivity estimation of approximate string queries [10, 12, 13] and approximate entity extraction [1, 4, 22, 15].

## 8. CONCLUSION

In this paper, we have studied the problem of string similarity joins with edit-distance constraints. We propose a partition-based method to do efficient similarity joins. We first sort strings and then visit strings in order. We build inverted indices for the visited strings. For each string, we select some of its substrings and utilize the selected substrings to find similar string pairs using the inverted indices. We develop a position-aware method and a multi-match-aware method to select substrings. We prove that the multi-match-aware selection method can minimize the number of selected substrings. We also develop efficient techniques to verify candidate pair based on length difference. We propose an extension-based method and share the computations on common prefixes to further improve the verification performance. Experiments show that our method outperforms state-of-the-art studies on both short strings and long strings.

## 9. ACKNOWLEDGEMENT

This work was partly supported by the National Natural Science Foundation of China under Grant No. 61003004 and 60873065, the National Grand Fundamental Research 973 Program of China under Grant No. 2011CB302206, National S&T Major Project of China under Grant No. 2011ZX01042-001-002, and the "NExT Research Center" funded by MDA, Singapore, under the Grant No. WBS:R-252-300-001-490.